\documentclass[aps,preprint,prd,nofootinbib]{revtex4-1}
\usepackage{graphicx}
\usepackage{psfrag}
\usepackage{subfigure}
\usepackage{array}
\usepackage{graphicx}
\usepackage{amsmath}
\usepackage{amssymb}
\usepackage{mathrsfs}
\usepackage{bm}
\usepackage{slashed}
\usepackage{threeparttable}
\usepackage{multirow}
\usepackage[colorlinks, linkcolor=black, anchorcolor=black, citecolor=black]{hyperref}
\usepackage[amsmath,thmmarks,hyperref]{ntheorem}
\usepackage{appendix}
\usepackage{soul}
\usepackage{hyperref}
\usepackage{cleveref}
\usepackage{float}
\usepackage{threeparttable}

\allowdisplaybreaks[4]

\begin{document}

\title{Study of the weak radiative decay $B_c^+\to B_u^{\ast+}\gamma$}
\author{Jia Zhao$^1$\footnote{24s011006@stu.hit.edu.cn}, Tianhong Wang$^1$\footnote{thwang@hit.edu.cn}, Xi-Yu Ni$^1$\footnote{26B911023@stu.hit.edu.cn}, Zhi-Hui Wang$^2$\footnote{zhwang@nmu.edu.cn},  and Guo-Li Wang$^{3}$\footnote{wgl@hbu.edu.cn}\\}
\address{$^1$School of Physics, Harbin Institute of Technology, Harbin, 150001, China\\
$^2$School of Electrical and Information Engineering, North Minzu University, Yinchuan 750021, China\\
$^3$Department of Physics, Hebei University, Baoding 071002, China}

\baselineskip=20pt

\begin{abstract}

We investigate the rare radiative decay $B_c \to B_u^\ast\gamma$ within the Standard Model by using the instantaneous Bethe-Salpeter (BS) approach. The short distance (SD) $c \to u\gamma$ penguin, weak annihilation, and long distance (LD) vector meson dominance (VMD) contributions are all included. In the calculation of LD contribution, we use two scenarios, namely the decay width of $\rho$ being set to zero or kept as the experimental value, which gives $\mathrm{Br}_{\rm LD}=5.0\times10^{-9}$ and $7.3\times10^{-8}$, respectively, indicating an important role of $\Gamma_\rho$ in the calculation. The SD contribution is comparable in magnitude to the LD contributions, but one order larger than the annihilation contribution. This shows the $B_c \to B_u^\ast\gamma$ channel can provide an important way to study the flavor changing neutral current processes of charm quark. 

\end{abstract}

\maketitle

\section{Introduction}

In the Standard Model (SM), flavor changing neutral current (FCNC) decays of charm quark are forbidden at tree level, and can only proceed via loop diagrams~\cite{Buchalla:1995}, making rare charm decays sensitive probes of short-distance dynamics and possible new physics (NP).  Experimentally, the weak radiative decays of charmed mesons and baryons have been widely studied~\cite{Gisbert:2020vjx}. BESIII set an upper limit of $10^{-5}$ level to the branching ratios of $D^{+} \to \gamma \rho^{+}$ and $D^{+} \to \gamma K^{\ast+}$~\cite{BESIII:2024zfv}; a search for $D_s^+ \to \rho(770)^+\gamma $ yielded a hint of signal with a branching fraction of $2.2 \times 10^{-4}$~\cite{BESIII:2024thr}. For the neutral charmed meson, Belle has measured the branching ratios of weak radiative decays and corresponding CP violation~\cite{Belle:2016mtj}. In the baryon sector, Belle~\cite{Belle:2022raw} and BESIII~\cite{BESIII:2022aok} have made searches for the weak radiative decays $\Lambda_c^+ \to \Sigma^+ \gamma$ and $\Xi_c^0 \to \Xi^0 \gamma$, and provide upper limits on the branching fractions. Furthermore, in the context of new physics searches, BESIII has explored the massless dark photon through the FCNC processes $D^0 \to \omega \gamma'$ and $D^0 \to \gamma \gamma'$, providing stringent constraints on the $c\to u\gamma'$ coupling \cite{BESIII:2024rkp}. Besides the charmed hadrons, BESIII also performed measurements for the FCNC processes of charmonia~\cite{Li:2026,BESIII:2017pez,BESIII:2024kkx,BESIII:2025wsu}, such as $J/\psi\to D^0l^+l^-$, $\psi(3686)\to D^0l^+l^-$, and $J/\psi\to\gamma D^0$, and set upper limits on these rare processes.

On the theoretical side, the FCNC decays of charm quark have been studied by different methods. For the charmed mesons, Refs.~\cite{deBoer:2017,Adolph:2018} analyzed radiative rare decays and considered the contributions of SM and different NP models. For charmed baryons, the branching fraction of $\Lambda_c^+ \to \Sigma^+ \gamma$ was calculated in Ref.~\cite{Shi:2022} by using light-cone sum rules, yielding a value of $(1.03 \pm 0.36) \times 10^{-4}$, which is consistent with the experimental bounds. Ref.~\cite{Zhang:2024} studied the rare decay mode $\Lambda_c^+ \to pl^+l^-$, and presented the constraint to the NP models. Within the framework of QCD sum rules, Ref.~\cite{Wang:2009} calculated the branching ratio of $J/\psi\to \bar D^{(*)0} l^+ l^-$, and the result is at $10^{-13}$ level, which is much smaller than the current experimental upper bound. In the theoretical study of charmed hadron decays, short distance (SD) signals are obscured by long distance backgrounds, making it difficult to distinguish the effects of new physics. How to extract information about the SD dynamics remains a longstanding challenge in rare charm decays \cite{Gisbert:2024,Fajfer:2001sa}.

Besides charmed mesons and baryons, the $B_c$ meson, as a two-heavy-flavor-quark bound state, can also provide a distinctive laboratory for the study of heavy quark flavor physics. Experimentally, almost all existing results of $B_c$ decay channels are associated with the $b$ quark decays, while for the $c$ quark sector, only the nonleptonic decay mode $B_c^+\to B_s^0\pi^+$ has been observed~\cite{LHCb:2013}. With the full Run 3 dataset and future Run 4 (HL-LHC) data-taking, LHCb will accumulate vastly larger $B_c$ samples~\cite{Gao:2010}, and more decay channels, even the FCNC processes, are expected to be observed. Theoretically, the weak decay properties of the $B_c$ meson have been extensively studied by using various approaches~\cite{Chang:1992,Kiselev:2000,Ivanov:2006}. As to the rare radiative decay mode $B_c\to B_u^\ast \gamma$, relevant research is carried out in Refs.~\cite{Fajfer:1999,Aliev:1999,Losacco:2024}. However, calculations from different models exhibit significant discrepancies. Unlike rare charm-meson and charm-baryon decays, where the LD effects generally dominate, in the $B_c$ case, the SD and LD contributions are comparable, which may facilitate the investigation of $c$-quark FCNC decays. Therefore, it is necessary to apply different methods to perform further studies of this decay process, such as precise calculations of the hadronic matrix elements and a more refined analysis of the contributions from different Feynman diagrams.

The paper is organized as follows. In the second part, we introduce the effective weak Hamiltonian for the $c \to u$ transition process, along with the transition amplitudes from short distance, weak annihilation and long distance contributions, and then calculate the hadronic transition matrix elements for the relevant processes by using the Bethe-Salpeter (BS) method. In the third part, we discuss the numerical results for the form factors and decay width of $B_c \to B^\ast_u\gamma$. Finally, we present a summary in the fourth part.

%%%%%%%%%%%%%%%%%%%%%%%%%%%%%%%%%%%%%%%%%%%%%%%%%%%%%%%

\section{Theoretical Formalism}
\label{sec2}
\begin{figure}[htbp]
	\centering
	\graphicspath{{figures/}}
	\subfigure[]{
		\includegraphics[width=2.5in,height=1.5in]{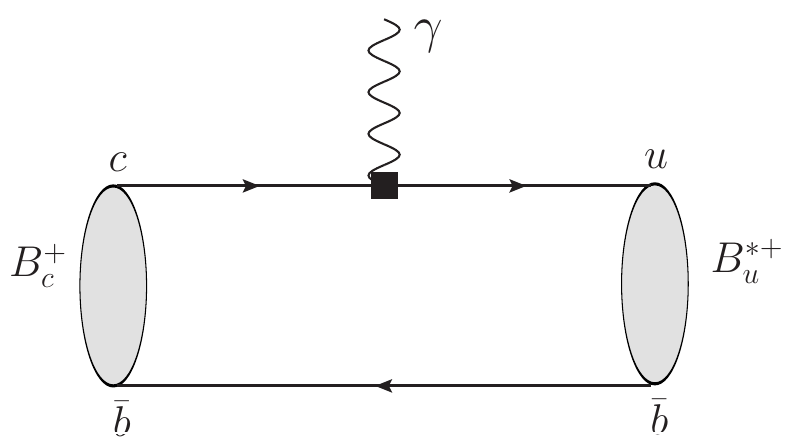}\label{1a}
	}
	\quad
	\subfigure[]{
		\includegraphics[width=2.7in,height=1.3in]{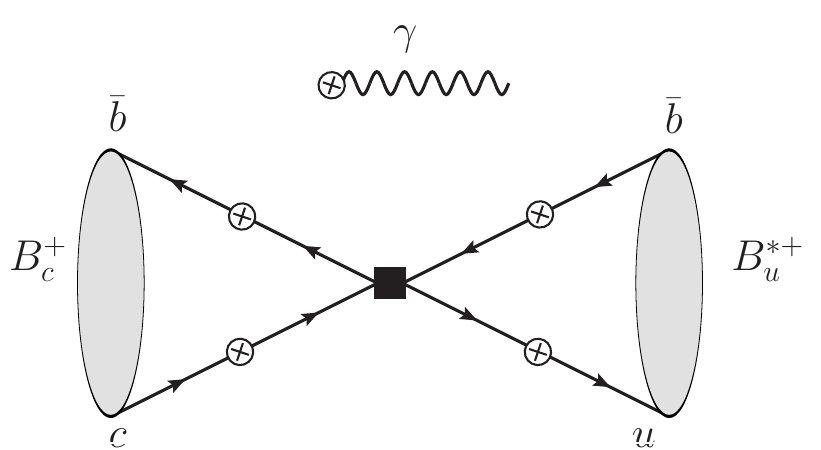}\label{1b}
	}
	\quad
	\subfigure[]{
		\includegraphics[width=2.5in,height=1.6in]{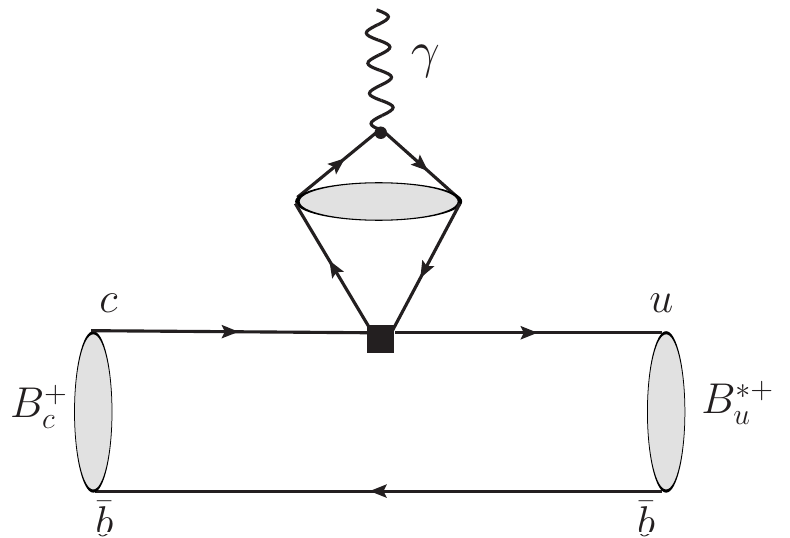}\label{1c}
	}
	\caption{Feynman diagrams of $B_c^+ \rightarrow B_u^{\ast+} \gamma$. The penguin diagram is shown in (a). The annihilation diagram is shown in (b), where the photon can be emitted from
		each quark line, denoted by $\otimes$. The long-distance contribution is shown in (c) with the intermediate meson being $\rho$, $\omega$, or $\phi$.}
	\label{fig1}
\end{figure}

The $B_c\to B_u^\ast \gamma$ decay amplitude is decomposed into three contributions: the SD penguin (Pen) contribution shown in Fig.~\ref{fig1}(a), the weak annihilation (Ann) contribution in Fig.~\ref{fig1}(b), and the LD vector meson dominance (VMD) contribution in Fig.~\ref{fig1}(c). The relevant $c\to u$ effective Hamiltonian is written as \cite{Greub:1996,Fajfer:2003}
\begin{equation}
	\begin{aligned}
	&H_{\mathrm{eff}}\left(M_{W}>\mu>m_{b}\right)=\frac{ 4G_{F}}{\sqrt{2}} \sum_{q=d, s, b} V_{c q} V_{u q}^\ast\left[C_{1}(\mu) O_{1}^{(q)}+C_{2}(\mu) O_{2}^{(q)}\right],\\
	&H_{\mathrm{eff}}\left(m_{b}>\mu>m_{c}\right)  =\frac{ 4G_{F}}{\sqrt{2}} \sum_{q=d, s} V_{c q} V_{u q}^\ast\left[C_{1}(\mu) O_{1}^{(q)}+C_{2}(\mu) O_{2}^{(q)}+\sum_{i=3}^{8} C_{i}(\mu) O_{i}\right],
	\end{aligned}
\end{equation} 
where $C_i$s are the Wilson coefficients which depend on the energy scale $\mu$. $V_{cq}$ and $V_{uq}$ are the relevant Cabibbo-Kobayashi-Maskawa (CKM) matrix elements, with $q$ taking $d$ or $s$ at scale $\mu \simeq m_c$.  The effective operators $O_i$s are defined as
\begin{equation}
	\begin{aligned}
	&O_{1}^{(q)}=(\bar{u}_{L}^{\alpha} \gamma_{\mu} q_{L}^{\beta})(\bar{q}_{L}^{\beta} \gamma^{\mu} c_{L}^{\alpha}),\\
	&O_{2}^{(q)}=(\bar{u}_{L}^{\alpha} \gamma_{\mu} q_{L}^{\alpha})(\bar{q}_{L}^{\beta} \gamma^{\mu} c_{L}^{\beta}),\\
	&O_{7}=\frac{e}{16 \pi^{2}}\bar{u} \sigma_{\mu \nu}m_{c}P_R c F^{\mu \nu}.
	\end{aligned}
\end{equation}
Here $O_7$ is referred to as magnetic penguin operator. In the expression, $P_R = \frac{1}{2}(1 + \gamma_{5})$ is the right-hand projection operators; $\alpha$ and $\beta$ are color indices; $F_{\mu\nu}$ is the electromagnetic field strength tensor. The expressions for the remaining operators can be found in in Ref.~\cite{Burd:2002} and are not listed explicitly here; their SD effects relevant to the radiative amplitude are incorporated into $C_7^{\rm eff}$.  

\subsection{The short distance penguin contribution}
The short distance contribution to $B_c^+ \rightarrow B_u^{\ast+} \gamma$ is induced by the FCNC transition $c \rightarrow u \gamma$. Such processes are particularly sensitive to electromagnetic dipole interactions and have therefore been proposed as useful probes of CP violation and possible new physics~\cite{Isidori:2012,deBoer:2017,deBoer:2018zhz,Biswas:2017eyn}. In SM, the electromagnetic dipole operator $O_7$ provides the dominant contribution.  The corresponding amplitude can be written as
\begin{equation}
	\mathcal{M}_{Pen}= \frac{G_{F}e}{4 \sqrt{2} {\pi}^2} V_{c s} V_{u s}^\ast C_{7}^{\mathrm{eff}} W^{\mu}_{T}{\epsilon }_{\mu}^{\ast}.
\end{equation}
In the above equation, ${\epsilon }_\mu$ is the polarization vector of the photon. $W^{\mu}_{T}$ is the hadronic matrix element, where the subscript $T$ indicates that only transverse photons contribute. With the effective operator $O_7$, $W_T^\mu$ can be written as 
\begin{equation}
	W^{\mu}_{T}=\langle B_{u}^\ast| \bar{u} i \sigma^{\mu \nu} k_{\nu} m_{c} (1+\gamma_{5}) c|B_{c}\rangle,
\end{equation}
where $k$ is the momentum of photon. For the $c\to u\gamma$ transition, we work at the characteristic scale $\mu\simeq m_c$ and adopt the two-loop QCD result for the effective Wilson coefficient from Ref.~\cite{Greub:1996}, namely $C_7^{\mathrm{eff}}(m_c)=-0.0068-0.020i$.

\subsection{The weak annihilation contribution}

The weak-annihilation contribution is governed by the charged-current four-quark transition $c\bar b\to u\bar b$. The relevant effective Hamiltonian has the form \cite{Fajfer:1999,Bauer:1986,Ju:2016}
\begin{equation}
	H_{\mathrm{eff}}^{Ann}=\frac{G_{F}}{\sqrt{2}} V_{c b} V_{u b}^{*} a_{1}^{\mathrm{eff}}(\mu)\left[\bar{u} \gamma^{\alpha}\left(1-\gamma_{5}\right) b\right]\left[\bar{b} \gamma_{\alpha}\left(1-\gamma_{5}\right) c\right],
\end{equation}
where $a_1^{\mathrm{eff}}$ is the effective Wilson coefficient. The renormalization scale in this case is taken to be $m_b$. Following Ref.~\cite{Fajfer:1999}, we adopt $a_1^{\mathrm{eff}}(m_b)=1.1$. The amplitude for the annihilation process can be written as
\begin{equation}
	\mathcal{M}_{Ann}= \frac{eG_{F}}{\sqrt{2}} a_{1}^{\mathrm{eff}} V_{c b} V_{u b}^{*} W_{A n n}^{\mu} \epsilon_{ \mu}^\ast,
\end{equation}
 with the hadronic matrix element
\begin{equation}
	W_{Ann}^{\mu}=W_{1}^{\mu}+W_{2}^{\mu}+W_{3}^{\mu}+W_{4}^{\mu}.
\end{equation}
Here, $W_{1,2,3,4}^\mu$ correspond to photon emission from the initial state $\bar b$ and $c$ quarks and the final state $\bar b$ and $u$ quarks, respectively, as shown in Fig.~\ref{1b}. Under the naive factorization scheme, they are given by
\begin{equation}
	\begin{aligned}
	W_1^{\mu}&=\langle B_{u}^\ast|\bar{u}\gamma_{\alpha}(1-\gamma_5)b|0\rangle\langle0|\bar{b}\left(-\frac{1}{3}\right)\gamma^{\mu}\frac{1}{\slashed p _{q_1}-m_{q_1}+i\epsilon}\gamma^{\alpha}(1-\gamma_5)c|B_{c}\rangle,\\
	W_2^{\mu}&=\langle B_u^\ast|\bar{u}\gamma_\alpha(1-\gamma_5)b|0\rangle\langle0|\bar{b}\gamma^\alpha(1-\gamma_5)\frac{1}{\slashed p _{q_2}-m_{q_2}+i\epsilon}\left(\frac{2}{3}\right)\gamma^\mu c|B_c\rangle,\\
	W_3^{\mu}&=\langle B_{u}^\ast|\bar{u}\gamma_\alpha(1-\gamma_5)\frac{1}{\slashed p _{q_3}-m_{q_3}+i\epsilon}\left(-\frac{1}{3}\right)\gamma^\mu b|0\rangle\langle0|\bar{b}\gamma^\alpha(1-\gamma_5)c|B_c\rangle,\\
	W_4^{\mu}&=\langle 
	B_{u}^\ast|\bar{u}\left(\frac{2}{3}\right)\gamma^\mu\frac{1}{\slashed p _{q_4}-m_{q_4}+i\epsilon}\gamma_\alpha(1-\gamma_5) b|0\rangle\langle0|\bar{b}\gamma^\alpha(1-\gamma_5)c|B_c\rangle.\label{W1234}
	\end{aligned}
\end{equation}
In the above equation, $p_{q_{(1-4)}}$ and $m_{q_{(1-4)}}$ denote respectively the momenta and masses of the intermediate quarks in the corresponding propagators.

\subsection{The long distance VMD contribution}

In this case, a vector meson $V$ is produced as the intermediate state, which then decays to a photon. Two kinds of processes, namely the color suppressed and color favored ones, are involved, which originate from the operators $O_1$ and $O_2$, respectively. The color-favored contribution comes from a heavy quarkonium as the intermediate state, which is highly suppressed by the large mass of the vector meson. Hence, we only consider the color suppressed process $B_c \rightarrow B^\ast_u V\rightarrow B^\ast_u \gamma$, where the vector meson $V$ can take $\rho,~\omega$, or $\phi$. The corresponding transition amplitude is decomposed into the product of three parts~\cite{Keum:2004},
\begin{equation}\label{vmd}
	\mathcal M_{\rm LD}=\sum_{V}\langle\gamma| ieA_{\nu} J_{e m}^{\nu}|V\rangle \frac{ie^{i\varphi_V}}{q_{V}^{2}-m_{V}^{2}+im_V\Gamma_V}\left\langle V B_u^{*}| H_{\mathrm{eff}}|B_c\right\rangle .
\end{equation}

In the above equation, the first term on the right represents the transition of the intermediate vector meson $V$ to a photon, where $A_\nu$ is the photon field and $J_{\text{em}}^\nu=-\sum_{q}Q_q\bar{q}\gamma^{\nu}q$ is the electromagnetic current with $Q_q$ being the electric charge of quark $q$. This matrix element can be expressed as
\begin{equation}
	\langle\gamma| ieA_{\nu} J_{e m}^{\nu}|V\rangle=-iQe~m_Vf_{V}\epsilon_V^{\nu}\epsilon_\nu^\ast,
\end{equation}
were $m_V$, $\epsilon_V$, and $f_V$ are the mass, polarization vector, and decay constant of the vector meson, respectively. $ Q $ is the effective charge which takes the following values~\cite{Keum:2004}: $ Q = (Q_u - Q_d) / \sqrt{2} $ for $ V = \rho $; $ Q = (Q_u + Q_d) / \sqrt{2}$ for $ V = \omega $; $ Q = Q_s $ for $ V = \phi$. Here we have applied the flavor wave functions of vector mesons.

The second term is the propagator of the intermediate vector meson with $q_V$ and $\Gamma_V$ being its momentum and total width, respectively. Generally, a strong phases $\varphi_V$ can be introduced for each vector meson. Here we will set these additional phases to zero~\cite{ODon:1991} for simplicity. As the photon is real, we have $q_V^2=k^2=0$, and the propagators become constants. The imaginary part can also produce a strong phase. However, it can be neglected when $V=\omega$ or $\phi$, as $\Gamma_\omega/m_\omega$ and $\Gamma_\phi/m_\phi$ are very small. As for the $\rho$ meson, it has a wide decay width, with $\Gamma_\rho/m_\rho\simeq 0.19$, corresponding to a strong phase of $11^\circ$ ~\cite{Keum:2004}. Therefore,  we retain this term to examine its impact on the final results.

The third term is the amplitude of $B_c \rightarrow B^\ast_u V$, which can be naively factorized as \cite{Fajfer:1998}
\begin{equation}
	\langle V B_u^{\ast}| H_{\mathrm{eff}}|B_c\rangle=  \frac{G_{F}}{\sqrt{2}} V_{c q} V_{u q}^\ast a_{2}^{\mathrm{eff}}\left\langle V| \bar{q} \gamma^{\mu}(1-\gamma^{5}) q|0\right\rangle\langle B^\ast_u| \bar{u} \gamma_{\mu}(1-\gamma^{5}) c|B_c\rangle,
\end{equation}
where $a_2^{\rm eff}$ is the effective Wilson coefficient, which is taken to be $a_2^{\rm eff}\simeq-0.5$ \cite{Fajfer:1999,Bauer:1986} at the renormalization scale $m_c$. In the above equation, the first matrix element on the right can be written as
\begin{equation}
	\left\langle V| \bar{q}\gamma^{\mu}(1-\gamma^5) q|0\right\rangle \equiv m_Vf_{V}\epsilon_V^{\ast\mu}.
\end{equation}
The decay constants in Eq.~(10) and (12) are defined at the energy scale $ q_V^2 = m_V^2 $. However, since the intermediate vector meson is off-shell, it is necessary to extrapolate these decay constants from $ q_V^2 = m_V^2 $ to $ q_V^2 = 0 $. We parameterize the off-shell extrapolation as $f_V(0)=\kappa_V f_V(m_V^2)$. For light vector mesons, one can assume $\kappa_V=1$~\cite{Keum:2004}.

Combining the above results, the amplitude for the cascade decay $ B_c \to B_u^\ast V \to B_u^\ast \gamma $ is 
\begin{equation}
	\mathcal{M}_{\rm LD}=  \frac{eG_{F}}{\sqrt{2}} V_{c s} V_{us}^\ast a_{2}^{\mathrm{eff}}C_{\rm VMD} W^\mu_{\rm LD}\epsilon^\ast_\mu.
\end{equation}
In the above equation, we have made the approximation that $V_{cs} V_{us}^\ast \simeq -V_{cd} V_{ud}^\ast$. $C_{\rm VMD}$ is defined as
\begin{equation}
	C_{\rm VMD}=\frac{f_{\rho}^2}{2(1-i\Gamma_{\rho}/m_{\rho})}-\frac{f_{\omega}^2}{6}-\frac{f_{\phi}^2}{3}.
\end{equation}
If $\Gamma_\rho$ is also taken to be zero, $C_{\rm VMD}$ will reduce to $f_\rho^2/2-f_\omega^2/6-f_\phi^2/3$, of which the three terms largely cancel each other by numerical calculation. This indicates the inclusion of $\Gamma_\rho/m_\rho$ term may have a sizable impact on the result. $W^\mu_{\rm LD}$ represents the hadronic transition matrix element $\langle B_u^\ast|\bar u\gamma_\mu(1-\gamma^5)c|B_c\rangle$. As the photon is on-shell, only the transverse part of the weak current contributes to the decay amplitude. Following the prescription adopted in Refs.~\cite{Fajfer:1997,Singer:1996}, by using the equation of motion of quarks and assuming the charm quark in the $B_c$ meson being at rest, we get
\begin{equation}
	\epsilon^\ast_\mu W^\mu_{\rm LD}=\langle B_u^\ast| \bar{u} \frac{-i\sigma^{\mu\nu}(1+\gamma^{5}) k_{\nu}\epsilon^\ast_\mu}{m_c}c|B_c\rangle=-\frac{\epsilon^\ast_\mu W^\mu_T}{m_c^2}.
\end{equation}

\subsection{The Bethe-Salpeter method}

The BS equation provides a relativistic field description of two-body bound states and has been extensively employed to study heavy mesons, such as their spectroscopy as well as decay properties~\cite{wang:2022cxy,Geng:2019,Ju:2013}. In practical calculations, by using the instantaneous approximation for the interaction kernel, one can reduce the four-dimensional BS equation to the three-dimensional Salpeter equations. By solving these equations numerically, one can get the Salpeter wave functions of heavy mesons with different quantum numbers, which are applied to calculate the transition matrix elements. 

In the present work, we employ the instantaneous BS approach to describe both the $B_c$ and $B_u^*$ mesons and to evaluate the hadronic matrix elements relevant to the $B_c\to B_u^\ast\gamma$ decay. Within this framework, the wave functions for the $B_c$ meson ($J^P=0^-$) has the following form~\cite{Kim:2003}
\begin{equation}
	\varphi_i(q_{\perp})=M\left[\frac{\slashed P}{M} g_{1}(q_{\perp})+g_{2}(q_{\perp})+\frac{\slashed q_{\perp}}{M} g_{3}(q_{\perp})+\frac{\slashed P \slashed q_{\perp}}{M^{2}} g_{4}(q_{\perp})\right]~\gamma^{5},\label{15}
\end{equation} 
and for the $B^\ast$ meson ($J^P=1^-$), the wave function is~\cite{Wang:2005}
\begin{equation}
	\begin{aligned}
		\varphi_f(q_{f\perp})=&(q_{f\perp} \cdot \epsilon_f)  {\left[f_{1}(q_{f\perp})+\frac{\slashed P_f}{M_f} f_{2}(q_{f\perp})+\frac{\slashed q_{f\perp}}{M_f} f_{3}(q_{f\perp})+\frac{\slashed P_f \slashed q_{f\perp}}{M_f^{2}} f_{4}(q_{f\perp})\right] } \\
		& +M_f \slashed \epsilon_f\left[f_{5}(q_{f\perp})+\frac{\slashed P_f}{M_f} f_{6}(q_{f\perp})+\frac{\slashed q_{f\perp}}{M_f} f_{7}(q_{f\perp})+\frac{\slashed P_f \slashed q_{f\perp}}{M_f^{2}} f_{8}(q_{f\perp})\right].\label{16}
	\end{aligned}
\end{equation}
In the above two equations, $P_{(f)}$ and $M_{(f)}$ are the momentum and mass of the initial (final) meson, respectively. $\epsilon_f$ is the polarization vector of the final meson. $q_{(f)\perp}$ is defined as $q_{(f)\perp}\equiv q^\mu_{(f)}- \frac{P_{(f)} \cdot q_{(f)} }{M_{(f)}^2} P_{(f)}$, with $q_{(f)}$ being the relative momentum between the quark and antiquark in the initial (final) meson. $g_i$ and $f_i$ are radial wave fucntions, the eigen equations fulfilled by which can be found in Refs.~\cite{Kim:2003}.

For the penguin and LD contributions (see Eq.~(4) and (15)), the hadronic transition matrix element $W_{\mu}^{T}$ is included, which can be written as the overlap integral of the wave functions of two mesons,
\begin{equation}
	W_{\mu}^{T}=\int \frac{d^{3}q_\perp}{(2 \pi)^{3}} {\rm Tr}\left\{\frac{ \slashed P}{M} \bar{\varphi}_{f}^{++}(q_{f\perp}) i \sigma_{\mu \nu}(P-P_{f})^{\nu} m_{c} (1+\gamma_{5}) \varphi_{i}^{++}(q_\perp)\right\}.
\end{equation}
Here $\bar\varphi$ is defined as $\gamma^0\varphi^\dagger\gamma^0$. $\varphi^{++}$ is the positive-energy wave function, which is defined as $\Lambda_1^+\frac{\slashed P}{M}\varphi\frac{\slashed P}{M}\Lambda_2^+$ with $\Lambda_{1,2}^+$ being the positive-energy projector operators~\cite{Kim:2003}. The tensor and pseudotensor transition matrix elements can be respectively  written as
\begin{equation}
	\begin{aligned}
	\epsilon^{\ast}_{\mu}\left\langle B_{u}^\ast\left(P_f, \epsilon_f\right)| \bar{u} i \sigma^{\mu \nu} k_{\nu} c|B_{c}(P)\right\rangle_{k^2=0}&=i \epsilon^{\mu \nu \alpha \beta}  {\epsilon^{\ast}_{\mu}\epsilon_f^{\ast}}_{\nu} P_{f\alpha} P_{\beta} F_0,\\
	\epsilon^{\ast}_{\mu}\left\langle B_{u}^\ast\left(P_f, \epsilon_f\right)| \bar{u} i \sigma^{\mu \nu} k_{\nu} \gamma_{5} c|B_{c}(P)\right\rangle_{k^2=0} &=[({m}^{2}_{{B}_{c}}-{m}^{2}_{B^\ast_{u}})\epsilon_f^\ast\cdot\epsilon^\ast-2 P\cdot\epsilon^\ast P\cdot\epsilon_f^{\ast}]F_{1},\label{W^T}
	\end{aligned}
\end{equation}
where $F_0$ and $F_{1}$ are the form factors, which are functions of $k^2=(P-P_f)^2$. Since the photon is on-shell, the form factors have taken the values at $k^2=0$, which can be prove to fulfill the relation $F_0(0)=2F_1(0)$. Furthermore, the gauge invariance requires that only the transverse components of the matrix elements contribute, while terms proportional to the photon momentum $k^\mu$ do not appear in the amplitude.  

For the weak-annihilation contribution, the four hadronic matrix elements in Eq.~(8) have the following expressions,
\begin{equation}\label{W11}
\begin{aligned}
		W_{1}^\mu &=-\frac{N_c}{3}\int \frac{d^{3} q_{f\perp}}{(2 \pi)^{3}}{\rm Tr}\left\{\bar{\varphi}_{f} \gamma_{\alpha}\left(1-\gamma_{5}\right)\right\} \times\int \frac{d^{3}q_\perp}{(2 \pi)^{3}} \frac{{\rm Tr}\left[ (\gamma^{\mu}\slashed{k}-2 p_{q_1}^{\mu})\gamma^{\alpha}(1-\gamma_{5})\varphi_{i}\right]} {-2p_{q_1}\cdot k+i \epsilon},\\
		W_{2}^\mu &=\frac{2N_c}{3}\int \frac{d^{3} q_{f\perp}}{(2 \pi)^{3}} {\rm Tr}\left\{\bar{\varphi}_{f}\gamma_{\alpha}\left(1-\gamma_{5}\right) \right\}\times\int \frac{d^{3}q_\perp}{(2 \pi)^{3}}\frac{{\rm Tr}\left[\gamma^{\alpha}(1-\gamma_{5}) (2 p_{q_2}^{\mu}-\slashed{k}\gamma^{\mu})\varphi_{i}\right]} {-2p_{q_2}\cdot k+i \epsilon},\\
	    W_{3}^{\mu} &=-\frac{N_c}{3}\int \frac{d^{3} q_{f\perp}}{(2 \pi)^{3}} \frac{{\rm Tr}\left[\bar\varphi_{f}\gamma^{\alpha}(1-\gamma_{5}) (-\slashed{k}\gamma^{\mu}-2 p_{q_3}^{\mu})\right]} {2p_{q_3}\cdot k+i \epsilon}\times\int \frac{d^{3}q_\perp}{(2 \pi)^{3}} {\rm Tr}\left\{ \gamma_{\alpha}\left(1-\gamma_{5}\right)\varphi_{i}\right\}, \\
		W_4^{\mu}&=\frac{2N_c}{3}\int \frac{d^{3}q_{f\perp}}{(2 \pi)^{3}} \frac{{\rm Tr}\left[\bar\varphi_{f} (2 p_{q_4}^{\mu}+\gamma^{\mu}\slashed{k})\gamma^{\alpha}(1-\gamma_{5})\right]} {2p_{q_4}\cdot k+i \epsilon}\times\int \frac{d^{3} q_\perp}{(2 \pi)^{3}} {\rm Tr}\left\{ \gamma_{\alpha}\left(1-\gamma_{5}\right)\varphi_{i}\right\},
\end{aligned}
\end{equation}
where $N_c=3$ is the number of colors; the momenta of quarks and antiquarks $p_{q_i}$ are expressed as 
\begin{equation}
\begin{aligned}
p_{q1}&\simeq \frac{m_1}{m_1+m_2}P-q_\perp,\qquad
p_{q2}\simeq \frac{m_2}{m_1+m_2}P+q_\perp,\\
p_{q3}&\simeq \frac{m_3}{m_3+m_4}P_f-q_{f\perp},\qquad
p_{q4}\simeq \frac{m_4}{m_3+m_4}P_f+q_{f\perp}.
\end{aligned}
\end{equation}

By using the following equations~\cite{Wang:2005,Cvetic:2004qg},
\begin{equation}
	\begin{aligned}
	\sqrt{N_c}\int \frac{d^{3} q_{f\perp}}{(2 \pi)^{3}} {\rm Tr}\left\{\bar{\varphi}_{f} \gamma_{\alpha}\left(1-\gamma_{5}\right)\right\}&=f_{B_u^\ast}M_f\epsilon_{f\alpha}^\ast,\\
	\sqrt{N_{c}} \int \frac{d^{3}q_\perp}{(2 \pi)^{3}} {\rm Tr}\left[\gamma^{\mu}\left(1-\gamma_{5}\right) \varphi_i\right]&=f_{B_c} P^{\mu},
	\end{aligned}
\end{equation}
where $f_{B_u^\ast}$ and $f_{B_c}$ are the decay constants of $B_u^\ast$ and $B_c$ mesons, respectively. The hadronic matrix element $W_{1,2,3,4}^{\mu}$ can be parameterized as
\begin{equation}
		W_i^\mu =\mathcal{N}_i\left[i\epsilon^{\mu\nu\alpha\beta}P_\nu P_{f\alpha}\epsilon^{*}_{f\beta}F^{(i)}_0+2P^\mu(\epsilon_f^*\cdot P)F^{(i)}_1+2P_1^\mu(\epsilon_f^*\cdot P)F^{(i)}_2+\left(m_{B_c}^2-m_{B_u^*}^2\right)\epsilon_f^{*\mu}F^{(i)}_3\right],
\end{equation}
where we have used the definitions $\mathcal{N}_{1}=-f_{B_{u}^{*}}M_{B_{u}^{*}}/\sqrt{3}$, $\mathcal{N}_{2}=2f_{B_{u}^{*}}M_{B_{u}^{*}}/\sqrt{3}$, $\mathcal{N}_{3}=-f_{B_{c}}/\sqrt{3}$  and $\mathcal{N}_{4}=2f_{B_{c}}/\sqrt{3}$. The form factors $F^{(i)}_{0,1,2,3}$ parameterize the second integrals in $W_{1,2}^\mu$ and the first ones in $W_{3,4}^\mu$. The gauge invariance requires that the annihilation amplitude fulfills the Ward identity, namely $k_\mu W^\mu_{\rm Ann}=0$. One can easily check that this is need the case by using Eq.~(\ref{W11}). 

%%%%%%%%%%%%%%%%%%%%%%%%%%%%%%%%%%%%%%%%%%%%%%%%%%%%%%%%%%

\section{Numerical Results}
In this section, we present the numerical results for the form factors and different contributions to the branching fraction of the $B_c\to B_u^\ast\gamma$ decay. In solving the BS equation under instantaneous approximation, we adopt the Cornell potential, which in the momentum space has the following form~\cite{Kim:2003},
\begin{equation}
	\begin{aligned}
		V({\vec q})&=V_s(\vec{q})
		+\gamma_0\otimes\gamma^0V_v(\vec{q}),\\
		V_{s}(\vec{q})
		&=-\left(\frac{\lambda}{\alpha}+V_0\right)\delta^{3}(\vec{q})
		+\frac{\lambda}{\pi^{2}}\frac{1}{(\vec{q}^{2}+\alpha^{2})^{2}},\\
		V_v(\vec{q})&=-\frac{2}{3\pi^{2}}
		\frac{\alpha_{s}(\vec{q})}{\vec{q}^{2}+\alpha^{2}},
	\end{aligned}
\end{equation}
with the running coupling constant expressed as
\begin{equation}
	\begin{aligned}
		\alpha_s(\vec{q})=\frac{12\pi}{27}
		\frac{1}{{\rm{ln}}\left(a+\frac{\vec{q}^2}{\Lambda^2_{QCD}}\right)},
	\end{aligned}
\end{equation}
where $a=e=2.71828$, $\alpha=0.06$ GeV, $\lambda=0.21$ ${\rm GeV}^2$, $\Lambda_{QCD}=0.27$ GeV. The masses of constituent quarks take the values~\cite{Ju:2013}: $m_b=4.96~\mathrm{GeV}$, $m_c=1.62~\mathrm{GeV}$, and $m_u=0.305~\mathrm{GeV}$. The masses and decay widths of mesons, and the CKM matrix elements are taken from the Particle Data Group~\cite{PDG:2024}.  The decay constants of heavy mesons are calculated in Eq.~(22), which are $f_{B_c}=-0.439~\mathrm{GeV}$ and $f_{B_u^\ast}=0.288~\mathrm{GeV}$. For the decay constants of light vector mesons, we adopt the results from Ref.~\cite{Bhar:2015}: $f_{\rho} = 0.216$~ GeV, $f_{\omega} = 0.197$~GeV, and $f_{\phi} = 0.233$~GeV. 

The numerical results of form factors are presented in Table~\ref{tab1}. For $W^T_\mu$, the two form factors fulfill the relation $F_1(0)=2F_2(0)$ as mentioned forward. In our calculation, we get $F_1(0)=1.15$, which is close to the prediction $F_1(0)=0.9$ of the QCD sum rules method~\cite{Aliev:1999}. In Ref.~\cite{Fajfer:1999}, the ISGW method is applied, and the result is 0.48, which is about one half of our result. For the annihilation process, there are four form factors, which exhibit a nontrivial dependence on the photon emission topology. One can see that the form factors of $W_{1,3}^\mu$ are real, while those of $W_{2,4}^\mu$ have imaginary parts. The reason is that for $W_{2,4}^\mu$ the quark propagator has poles for certain values of $q_\perp$ or $q_{f\perp}$.  Calculation shows that $W_{2,4}^\mu$ give the main contribution to the annihilation process. As the value of the imaginary part is comparable to that of the real one, it is important to include both parts. 
\begin{table}[htbp]
	\centering
	\setlength{\tabcolsep}{15pt}
	\caption{Form factors of the hadronic matrix elements 
		$W_\mu^T$ and $W_\mu^{1,2,3,4}$ at $k^2=0$.}
	\label{tab1}
	\begin{tabular}{c c c c c}
		\hline
		~ & $F_0$ & $F_1$ & $F_2$ & $F_3$ \\
		\hline
		$W_\mu^T$
		& $1.15$
		& $0.57$
		& $-$
		& $-$ \\
		
		$W_\mu^1$
		& $0.031$
		& $-0.0022$
		& $-0.011$
		& $-0.010$ \\
		
		$W_\mu^2$
		& $0.11-0.052i$
		& $0.0064+0.028i$
		& $-0.055-0.062i$
		& $0.072+0.034i$ \\
		
		$W_\mu^3$
		& $-0.069$
		& $-0.012$
		& $-0.039$
		& $-0.054$ \\
		
		$W_\mu^4$
		& $-0.51+0.40i$
		& $-0.67+0.39i$
		& $0.54-0.26i$
		& $0.21-0.13i$ \\
		\hline
	\end{tabular}
\end{table}

 Different contributions to the branching fraction are summarized in Table~\ref{tab2}. For the short distance contribution, our result is close to that in Ref.~\cite{Aliev:1999} and five times the value in Ref.~\cite{Fajfer:1999}. This is a direct result of the form factors. The annihilation contribution is one order of magnitude smaller than that of SD in our calculation. This is mainly because the small factor $V_{cb}V_{ub}^\ast$ appears in such process. The result in Ref.~\cite{Fajfer:1999} is a factor of 1.5 larger than ours. Although the numerical values are comparable, their adopted amplitude expression differs from ours and contains only $F_0$. The result of SD+Ann is slightly smaller than that of SD, which means two amplitudes interfere mildly destructively. 
 
 \begin{table}[htbp]
 	\centering
 	\setlength{\tabcolsep}{10pt}
 	\caption{Different contributions to the branching fraction of $B_c\to B_u^\ast \gamma$. The second column gives the results in the zero-width approximation. The third column gives the results including the width of $\rho$ meson, but assuming $\Gamma_\omega=\Gamma_\phi=0$.}
 	\label{tab2}
 	\begin{threeparttable}
 		\scalebox{1.0}{
 			\begin{tabular}{lcccc}
 				\hline
 				Contribution&This work ($\Gamma_V=0$)&This work ($\Gamma_{\rho}\neq 0$)&ISGW \cite{Fajfer:1999}\tnote{a}&QCDSR \cite{Aliev:1999}\\
 				\hline
 				$\mathrm{Br}_{\rm SD}$&$2.6\times 10^{-8}$&$2.6\times 10^{-8}$&$4.7 \times 10^{-9}$&$2.0 \times 10^{-8}$\\
 				$\mathrm{Br}_{\rm Ann}$&$2.1 \times 10^{-9}$&$2.1 \times 10^{-9}$&$3.2 \times 10^{-9}$&-\\
 				$\mathrm{Br}_{\rm SD+Ann}$&$2.5\times 10^{-8}$&$2.5\times 10^{-8}$&-&-\\
 				$\mathrm{Br}_{\rm LD}$&$5.0 \times 10^{-9}$&$7.3\times 10^{-8}$&$1.8 \times 10^{-9}$&-\\
 				$\mathrm{Br}_{\rm Total}$&$4.0\times 10^{-8}$&$5.2\times 10^{-8}$&$8.5 \times 10^{-9}$&-\\
 				\hline
 			\end{tabular}
 		}
 		\begin{tablenotes}
 			\footnotesize
 			\item[a] For the ISGW results, $\mathrm{Br}_{\rm Ann}=3.2\times10^{-9}$ and $\mathrm{Br}_{\rm LD}=1.8\times10^{-9}$ are derived, respectively, from the pole and VMD components of $A_{\rm PV}$ and $A_{\rm PC}$ reported in Ref.~\cite{Fajfer:1999}, where these two values are not presented.
 		\end{tablenotes}
 	\end{threeparttable}	
 \end{table}

For the long distance contribution, we consider two scenarios. First, we neglect the decay widths of three light vector mesons ($\Gamma_V=0$). As mentioned in Sec. II, the cancellation between the propagators leads to a small $C_{\rm VMD}$ in Eq.~(13), which further results in a small LD contribution. Combining all the contributions gives $\mathrm{Br}_{\rm Total}=4.0\times10^{-8}$, corresponding to a constructive interference between three amplitudes. Second, we keep $\Gamma_\rho$ but assume $\Gamma_\omega=\Gamma_\phi=0$ (see Eq.~(14)). $C_{\rm VMD}$ changes from $-1.24\times10^{-3}~{\rm GeV}^2$ in scenario one to $(-2.05+4.27i)\times10^{-3}~{\rm GeV}^2$, which leads to an increase of the LD contribution by one order of magnitude. This implies that the inclusion of meson widths is essential for the calculation. We should mention that, in this case, the real part of $C_{\rm VMD}$ is small, thereby highlighting the importance of the imaginary part. For other processes, such as $b\to s$, including the finite width will not significantly affect the calculation.  Unlike the result of LD, the total branching fraction does not vary much, which comes from the interference term turning from constructive into destructive. One can also include $\Gamma_\omega$ and $\Gamma_\phi$. Since they are much smaller than the corresponding masses, calculation shows ${\mathrm Br}_{\rm LD}$ changes only by $5.5\%$, which is negligible.

\section{Summary}
\label{Conc}
We have studied the rare radiative decay $B_c\to B_u^\ast\gamma$ within the Standard Model by using the instantaneous Bethe–Salpeter approach. This decay process has a different topology from charmed meson decays and can provide an additional complement to the study of charm-quark FCNC processes. The $c\to u\gamma$ penguin, weak annihilation, LD contributions and their interference effects are investigated. We obtain $\mathrm{Br}_{\rm SD}=2.6\times10^{-8}$, which is consistent with that of the QCD sum rules method but larger than the ISGW result. When computing the LD contributions, cancellations among light-meson propagators yield a result that is one-fifth of the SD contribution. However, once the $\rho$ meson width is included, the result increases by one order of magnitude. Although the LD contributions are strongly affected by the $\rho$ meson width, the total branching ratios obtained in the two scenarios are close to each other.

\acknowledgments

This work was supported by the National Natural Science Foundation of China (NSFC) under Grants Nos. 12375085, 12365013 and 12575097. Zhi-Hui Wang was supported by the Natural Science Foundation of Ningxia
(2023AAC03276).

\bibliographystyle{apsrev4-1}
\bibliography{ref}

\end{document}